\documentclass[prb,10pt,twocolumn]{revtex4-1}
\usepackage{amsmath}
\usepackage{amssymb}
\usepackage{graphicx}
\usepackage{hyperref}
\usepackage{framed}
\usepackage{comment}

\newcommand{\eqnref}[1]{Eq.~(\ref{#1})}
\newcommand{\ket}[1]{|#1\rangle}

\newcommand{\theapp}{Appendix \ref{sec_equivalence}}
\begin{document}
\title{Classification of topological phases in periodically driven interacting systems}
\author{Dominic V. Else}
\affiliation{Department of Physics, University of California, Santa Barbara, CA 93106, USA}
\author{Chetan Nayak}
\affiliation{Department of Physics, University of California, Santa Barbara, CA 93106, USA}
\affiliation{Microsoft Research, Station Q, Elings Hall, University of
California, Santa Barbara, CA 93106, USA}
\begin{abstract}
We consider topological phases in periodically driven (Floquet) systems exhibiting many-body localization, protected by a symmetry $G$. We argue for a general correspondence between such phases and topological phases of undriven systems protected by symmetry $\mathbb{Z} \rtimes G$, where the additional $\mathbb{Z}$ accounts for the discrete time translation symmetry.  Thus, for example, the bosonic phases in $d$ spatial dimensions without intrinsic topological order (SPT phases) are classified by the cohomology group $H^{d+1}(\mathbb{Z} \rtimes G, \mathrm{U}(1))$. For unitary symmetries, we interpret the additional resulting Floquet phases in terms of the lower-dimensional SPT phases that are pumped to the boundary during one time step. These results also imply the existence of novel symmetry-enriched topological (SET) orders protected solely by the periodicity of the drive.
\end{abstract}
\maketitle

There are now many known examples of phases of matter which are distinguished not by the symmetries they break spontaneously but through more subtle ``topological'' orders \cite{Wen1990}. Most such phases are not robust to thermal 
excitations and therefore were thought to exist only at zero temperature \cite{Nussinov2008,Alicki2009}. However, recently it has been appreciated that, in the presence of strong disorder, it is possible for highly excited eigenstates of a many-body system to be \emph{many-body localized (MBL)} \cite{Basko2006,Oganesyan2007,Pal2010a,Bardarson2012,Serbyn2013a,Serbyn2013b,Bauer2013,Imbrie2014a,Huse2014,Nandkishore2015}. Such MBL states are not thermal, and indeed more closely resemble gapped ground states; for example, they obey an area law for the entanglement entropy. This means that they can exhibit topological phases previously thought to be restricted to zero temperature\cite{Huse2013,Chandran2014,Slagle2015,Potter2015,Bahri2015}.

The lifting of the restriction to ground states also allows us to consider more general ``Floquet'' systems  \cite{Kitagawa2010,Lindner2011,Jiang2011,Thakurathi2013,Asboth2014,
Titum2014,Titum2015,Nathan2015,Iadecola2015,Carpentier2015}, in which the Hamiltonian $H(t)$ is allowed to vary in time, but with periodicity $T$. The ``eigenstates'' of such a system are the eigenstates of the Floquet operator $U = U(T)$ which describes the unitary evolution of the system over one time period. Such eigenstates can also be MBL in the presence of strong disorder \cite{DAlessio2013,Ponte2015a,Lazarides2015,Ponte2015,Abanin2014,Abanin2015}, and hence can exhibit topological phases. However, the classification of topological phases in such ``Floquet-MBL'' systems is in general richer than in the stationary case.

Recently, progress has begun to be made in understanding the classification of topological phases in Floquet-MBL systems with interactions\cite{Khemani2015a,VonKeyserlingk2016}. In particular, Ref.~\onlinecite{VonKeyserlingk2016} classified phases with a symmetry $G$ and no intrinsic topological order (i.e.\ \emph{symmetry-protected topological (SPT) phases} \cite{Gu2009,Pollmann2010,Pollmann2012,Fidkowski2010,Chen2010,Chen2011,Schuch2011a,Fidkowski2011, Chen2011b,Chen2013,Levin2012,Vishwanath2013,Wang2014, Kapustin2014,Gu2014,Else2014,Burnell2014,
Wang2015,Cheng2015}) in (1+1)-D. The purpose of this Communication is to re-express the classification of Ref.~\onlinecite{VonKeyserlingk2016} in a concise way, which we feel clarifies the issues involved and streamlines the derivation. We then consider natural extensions, building up to a (conjectured) general correspondence between topological phases in Floquet-MBL systems with symmetry group $G$, and topological phases in stationary systems with symmetry group $\mathbb{Z} \rtimes G$, where the extra $\mathbb{Z}$ accounts for the discrete time-translation symmetry.

\emph{Assumptions}.--
We will assume that the Floquet operator $U$ can be expressed as a time evolution of a local time-dependent Hamiltonian $H(t)$, with $H(t+T) = H(t)$. Thus,
\begin{equation}
\label{time_evolution}
U = \mathcal{T} \exp\left(-i\int_0^T H(t) dt\right), \quad \mbox{$\mathcal{T}$ = time-ordering}.
\end{equation}
where we assume that the Hamiltonian $H(t)$ is invariant under a representation $V(g)$ of a symmetry group $G$, where $G$ can contain anti-unitary elements corresponding to a time reversal symmetry. For anti-unitary $g \in G$, what we mean by the Hamiltonian ``being invariant'' is that $V(g) H(t) V(g)^{-1} = H(T-t)$. This ensures that, in general,
\begin{equation}
\label{repr_semidirect}
V(g) U V(g)^{-1} = U^{\alpha(g)},
\end{equation}
where $\alpha(g) = -1$ if $g$ is anti-unitary and $+1$ otherwise.

\emph{The SPT classification}.-- The classification of Ref.~\onlinecite{VonKeyserlingk2016} can be re-expressed in the following way. We define an enlarged  symmetry group $\widetilde{G}$ to be
the \emph{full} symmetry group of the system, \emph{including} the discrete time
translation symmetry inherent in the Floquet setup. Thus, if all of the
symmetries of $G$ are unitary, we have $\widetilde{G} = G \times \mathbb{Z}$.
More generally, for anti-unitary elements $g \in G$, we have $g \mathbb{T}
g^{-1} = \mathbb{T}^{-1}$, where $\mathbb{T}$ is the generator of time
translations. Thus, in general $\widetilde{G}$ is a semi-direct product $\widetilde{G} = \mathbb{Z} \rtimes G$. Then in the bosonic case, the classification of Ref.~\onlinecite{VonKeyserlingk2016} can be reformulated as follows (see \theapp{} for a proof):

\begin{framed}
\textbf{Result 1}.
The symmetry-protected topological phases in a periodically driven (1+1) bosonic system exhibiting MBL are classified by the second cohomology group $H^2(\widetilde{G}, U(1))$.
\end{framed}
(Here, and later, we will take it to be implicit that $U(1)$ is to be
interpreted as a non-trivial $\widetilde{G}$-module, with anti-unitary elements
of $\widetilde{G}$ acting as inversion, as in the original classification of SPT
phases with anti-unitary symmetries, e.g.\ see Ref.~\onlinecite{Chen2013}).

Recall that the bosonic topological phases in a stationary system are classified by $H^2(G, \mathrm{U}(1))$; to obtain the classification in a driven system one simply replaces $G$ by $\widetilde{G}$. In retrospect, this result should be quite natural. Indeed, the classification of stationary SPT phases in (1+1)-D \cite{Pollmann2010,Chen2011,Schuch2011a,Pollmann2012}, though sometimes expressed in terms of Hamiltonians, is really at its core a classification of \emph{short-range entangled} states (states which are equivalent to a product state by a local unitary) invariant under some local (anti-)unitary representation of a symmetry group (see Appendix \ref{deriving_spt} for more details). The gapped ground states of a Hamiltonian are examples of such states, but so are MBL eigenstates of a Floquet operator. (We could even consider eigenstates of the Floquet operator which are not MBL but are separated from all other eigenstates by a quasienergy gap). Thus, the standard classification of (1+1)-D SPT phases can be applied to any such states. However, there is one difference in the Floquet case: as well as the representation of the symmetry $G$, a Floquet eigenstate is, by definition, also invariant (up to a phase factor) under the Floquet operator $U$, which is a local unitary since it is the time evolution of a local Hamiltonian. Therefore, we should really include $U$ in the symmetry group to obtain the full classification. [\eqnref{repr_semidirect} ensures that we then have a representation of the enlarged symmetry group $\widetilde{G}= \mathbb{Z} \rtimes G$].

It is true that, when classifying SPT phases, one normally assumes that the action of the symmetry is ``on-site'', that is, that each symmetry operator $V(g)$ is a tensor product of its action on each site of the lattice, $V(g) = [v(g)]^{\otimes N}$, which would not be true of the Floquet unitary $U$. However, all we actually need is that all the symmetry operators (including the Floquet unitary $U$) can be \emph{restricted} to a region $A$ with boundary while still remaining a representation of $\widetilde{G}$, where by ``restriction'' of a local unitary $U$ we mean\cite{Else2014} a unitary $U_A$ acting only on the region $A$ which acts the same as $U$ in the interior of $A$, well away from the boundary. See Appendix \ref{deriving_spt} for the derivation of the classification, given such an assumption.

To see that such a restriction is possible, consider for simplicity the case of unitary symmetries. Then if the Hamiltonian $H(t)$ can be written as a sum $H(t) = \sum_X h_X(t)$ of terms supported on local regions $X$ [each of which commutes with the symmetry $V(g)$], then we can define the restriction of the Floquet operator by simply retaining only the terms which act within $A$, or in other words:
\begin{equation}
U_A = \mathcal{T} \exp\left(-i\int_0^T dt \sum_{X \subseteq A} h_X(t)\right).
\end{equation}
Meanwhile, we define the restriction of $V_A(g)$ in the obvious way, by only acting with the on-site action on sites contained within $A$. It is easily seen that $V_A(g)$ is still a representation of $G$, and $U_A$ commues with $V_A(g)$, so together they form a representation of $\widetilde{G} = \mathbb{Z} \times G$. Similar arguments can be made for anti-unitary symmetries.

We emphasize that our derivation of Result 1 is actually more general than that of Ref.~\onlinecite{VonKeyserlingk2016}. Firstly, in Ref.~\onlinecite{VonKeyserlingk2016} the result for non-Abelian $G$ was only stated as a conjecture. Our derivation clearly applies to such $G$ as well. Secondly, we did not need to assume, as did Ref.~\onlinecite{VonKeyserlingk2016} that \emph{all} the eigenstates of the Floquet operator are MBL; our classification result applies to any of the eigenstates that happen to be MBL, or separated from the rest of the quasienergy spectrum by a gap. Finally, since our derivation was based on individual eigenstates, it allows for the possibility of different SPT phases coexisting as eigenstates of a single Floquet operator, separated by an eigenstate transition \cite{Huse2013,Slagle2015}.

\emph{Higher dimensional results}.-- When stated in the form given here, classification result of Ref.~\onlinecite{VonKeyserlingk2016} has obvious generalizations to higher dimensions. In particular, in Ref.~\onlinecite{Else2014} we derived the classification of (2+1)-D SPT phases in ground states by considering how the symmetry acts on the boundary. In Ref.~\onlinecite{Else2014}, we did use the Hamiltonian to argue that the symmetry action on the boundary is well-defined; however, Appendix \ref{deriving_spt} shows how to formulate this concept for a single short-range entangled state without reference to a Hamiltonian (and without assuming that the symmetry in the bulk is on-site). Therefore, we can repeat the analysis of Ref.~\onlinecite{Else2014} (but taking care to include the Floquet unitary $U$ in the symmetry group), and one finds that

\begin{framed}
\textbf{Result 2}.
The symmetry-protected topological phases in a periodically driven (2+1)-D
bosonic system exhibiting MBL are classified by the third cohomology group $H^3(\widetilde{G}, U(1))$.
\end{framed}
Again, we simply replace $G \to \widetilde{G}$ compared to the usual stationary
case.
The anti-unitary case was not explicitly treated in Ref.~\onlinecite{Else2014}, but it is a straightforward generalization\cite{ElseUnpublished}. One can also prove a similar result for fermionic systems.

\emph{General correspondence between stationary and Floquet-MBL topological phases}.--
The above results relied on the method of Ref.~\onlinecite{Else2014}, which did not consider (at least, not in full generality) SPT phases in higher dimensions, or topological phases beyond SPT. Nevertheless, they
motivate us to formulate the following conjecture.
\begin{framed}
\textbf{Conjecture 1}.
The topological phases in a (bosonic/fermionic) periodically driven MBL system in $d$ spatial dimensions with on-site symmetry group $G$ are in one-to-one correspondence with the topological phases in a (bosonic/fermionic) \emph{stationary} MBL system in $d$ spatial dimensions with symmetry group $\widetilde{G} = \mathbb{Z} \rtimes G$ (as defined above).
\end{framed}
Here by ``topological phases'', we mean both symmetry-protected topological (SPT) phases and symmetry-\emph{enriched} topological (SET) phases \cite{Maciejko2010,Essin2013,Lu2013,Mesaros2013,Hung2013,Barkeshli2014,Cheng2015a}. The rationale for this conjecture is as follows. The classification of gapped ground states is known to depend only on the ground states themselves, not on their parent Hamiltonians \cite{Schuch2011a}. Furthermore, since eigenstates in an MBL system look, roughly speaking, like gapped ground states, one expects to obtain the same classification for such eigenstates. However, in a periodically driven system there is an extra local unitary, beyond the symmetries in the group $G$, under which these eigenstates are invariant (up to a phase factor) -- namely, the Floquet unitary $U$. Thus, one should treat $U$ as a symmetry for the purpose of obtaining the classification.

The only way we could envision this conjecture failing would be if the
non-on-site nature of the Floquet unitary $U$ turned out to be important, in a
way that it was not in the case of (1+1)-D and (2+1)-D SPT's. This seems to us
unlikely. In fact, we expect that \emph{any} derivation of the classification of SPT/SET phases -- or at least, any derivation which can be formulated in terms of short-range entangled states without reference to Hamiltonians -- could probably be applied just as well in the Floquet context, which would prove the conjecture.

We note, however, that probably not all topological phases which can
exist at zero temperature can be stabilized in MBL excited states
\cite{Potter2015}; for this reason, we have been careful to formulate Conjecture
1 in terms of a correspondence with stationary MBL systems, not with zero-temperature states.

\emph{Interpretation of the classification in terms of pumping.}--
Results 1 and 2, and Conjecture 1 in higher dimensions, imply that the classification of  SPT phases in bosonic Floquet-MBL systems in $d$ spatial dimensions is $H^{d+1}(\widetilde{G}, U(1))$. In the case of a unitary symmetry, such that $\widetilde{G}$ is just a direct product $\mathbb{Z} \times G$, we can give a simple physical interpretation of this result. From the K\"unneth formula for group cohomology \cite{Cheng2015a}, one finds that
\begin{equation}
\label{kunneth}
H^{d+1}(\mathbb{Z} \times G, \mathrm{U}(1)) = H^{d+1}(G, \mathrm{U}(1)) \times H^d(G, \mathrm{U}(1)).
\end{equation}
Thus, the classification is just the usual classification for ground states, plus an extra piece of data given by an element of $H^d(G, \mathrm{U}(1))$. We expect that this extra piece of data can be interpreted as characterizing the fact that each application of the Floquet unitary $U$ ``pumps'' an additional $(d-1)$-dimensional SPT phase onto the boundary. This is a generalization of the observation in Ref.~\onlinecite{VonKeyserlingk2016} that in (1+1)-D the extra data is the charge pumped onto each component of the boundary by the Floquet unitary.

A rough physical justification for this interpretation in (2+1)-D (which readily generalizes also to higher dimensions) is as follows. For simplicity we assume that $G$ is Abelian. One can then show that the $H^2(G, \mathrm{U}(1))$ piece of \eqnref{kunneth} can be extracted from a 3-cocycle $\omega(\widetilde{g}_1, \widetilde{g}_2, \widetilde{g}_3)$ of the full symmetry group $\widetilde{G}$ by calculating a 2-cocycle of $G$ according to
\begin{equation}
\label{dimreduction}
\omega(g_1, g_2) = \frac{\omega(\mathbb{T},g_1,g_2) \omega(g_1, g_2, \mathbb{T})}{\omega(g_1, \mathbb{T}, g_2)}.
\end{equation}
(where $\mathbb{T}$ is the generator of discrete time translations.)
The object \eqnref{dimreduction} has a familiar interpretation \cite{Bais1993}. Indeed, suppose we \emph{gauge} the full symmetry group $\widetilde{G} = \mathbb{Z} \times G$. Then the point excitations in the resulting twisted (2+1)-D gauge theory can be classified by the flux $\widetilde{g} \in \widetilde{G}$ they carry. In general, a particle carrying non-trivial flux also carries a \emph{projective} representation of the gauge group. In particular, \eqnref{dimreduction} describes the projective representation of the subgroup $G$ on a particle carrying flux $\mathbb{T}$. Now, in the original ungauged SPT phase, the analog of a flux is a ``symmetry twist defect'' \cite{Bombin2010,Barkeshli2013,Chen2015a,Barkeshli2014} which (since fluxes are confined) must occur at the endpoint of a symmetry twist line. The fact that the endpoints of such symmetry twist lines carry projective representations of $G$ (which can also be derived directly, using the theory of twist defects developed in Ref.~\onlinecite{Barkeshli2014}) shows that the lines themselves must be in a (1+1)-D SPT phase with respect to $G$. On the other hand, a \emph{closed} symmetry twist line (with no endpoints) on the boundary $\partial A$ of a region $A$ can be interpreted as the result of applying to the original MBL eigenstate the Floquet unitary $U$, restricted to the region $A$. The fact that such a state carries a (1+1)-D SPT on the boundary $\partial A$ indeed shows that the effect of $U$ is to pump a (1+1)-D SPT to the boundary.

On the other hand, we do not expect there to be any similarly simple physical picture in the anti-unitary case; in Ref.~\onlinecite{VonKeyserlingk2016} it was found that the extra data for (1+1)-D systems is a somewhat strange ``twisted'' representation of the symmetry with no obvious physical interpretation.

\emph{Floquet topological phases without symmetry}.--
The above considerations allow us to the establish the existence of topological
phases in driven MBL systems that are distinct in the Floquet context, even in
the absence of any additional symmetry, but not in the stationary case. Indeed,
imagine we take a Floquet system in (2+1) dimensions or higher, with symmetry group $\widetilde{G} = G \times
\mathbb{Z}$, and then gauge just the symmetry $G$. In general, gauging a subgroup of the full symmetry group relates SPT phases to
symmetry-enriched topological (SET) phases protected by the remaining global symmetry \cite{Mesaros2013,Hung2013,Barkeshli2014}; which, in this case, is simply the discrete
time translation symmetry.

\emph{Explicit realization}.-- We have already argued above that the invariants
which classify Floquet-MBL topological phases with symmetry $G$ should be the
same as in the case of stationary topological phases with symmetry $\mathbb{Z} \rtimes G$. However, one might ask whether there might be an
obstruction to realizing any of these ``potential''
Floquet-MBL topological phases in an explicit model. We argue that this is not
the case, \emph{provided} that the corresponding stationary topological phase with symmetry $\mathbb{Z} \rtimes G$ can
be realized in a stationary MBL system with symmetry $\widetilde{G}_n = \mathbb{Z}_n \rtimes G$ for some
sufficiently large $n$. Such a system, by definition, consists of a Hamiltonian $H$
which commutes with an on-site representation $V(\widetilde{g})$ of
$\widetilde{G}_n$. (A faithful on-site
representation of $\mathbb{Z}$ does not make sense in a lattice system with
finite-dimensional Hilbert space per site, hence why we consider $\mathbb{Z}_n$
instead. A system acted on by $\mathbb{Z}_n$ can always be thought of as being acted on by $\mathbb{Z}$ non-faithfully). Then we claim that the Floquet system with Floquet
operator $U = e^{i H T} V(\alpha)$ (where $\alpha$ is the generator of
$\mathbb{Z}_n$) indeed realizes the desired Floquet-MBL topological phase.

To see this, note 
 that the eigenstates of $H$ are also eigenstates of $V(\alpha)$ [since $H$ commutes with $V(\alpha)$  by assumption] and therefore of $U$. We can analyse the SPT order of these states by thinking of them either as eigenstates of a stationary system with symmetry $\widetilde{G}$, or as eigenstates of a Floquet system with symmetry $G$. In fact, the analysis proceeds identically in both cases, with only one difference: in the stationary context, the $\mathbb{Z}$ part of the symmetry is taken to be generated by $V(\alpha)$, whereas in the Floquet context, it is generated by $U$. However, we can make $U = V(\alpha)$ by sending $T \to 0$ continuously. Since the classification of topological phases is \emph{discrete}, we do not expect that this can change the diagnosed phase. This can be checked explicitly in the (2+1)-D SPT case.

\emph{Conclusion}.--
The perspective on topological phases in Floquet-MBL systems detailed in this
Communication opens up many intriguing questions for future study. Indeed, \emph{every}
phenomenon that has been studied in the usual stationary case -- for example,
symmetry fractionalization on topological excitations in symmetry-enriched
topological (SET) phases \cite{Essin2013,Barkeshli2014,Cheng2015} 
-- ought to have analogs in the Floquet-MBL case, but
in many cases the possibilities will be richer due to the extra $\mathbb{Z}$
symmetry. We leave further exploration of these phases and their physical
properties for future work.

\emph{Note added}.--
Soon after we posted this work on the arXiv, two more preprints appeared\cite{Potter2016,Roy2016} whose
results overlap with ours.

\begin{acknowledgments} We acknowledge support from the Microsoft
  Corporation. We thank C.~W.~von Keyserlingk for drawing our attention to this
  problem and providing us a copy of
  Ref.~\onlinecite{VonKeyserlingk2016} prior to its appearance on the arXiv. We thank the anonymous referees for encouraging us to clarify our arguments.
\end{acknowledgments}

\section{Appendix: Equivalence to the classification of von Keyserlingk and Sondhi}
\label{sec_equivalence}
%

Here we will show that the classification of Ref.~\onlinecite{VonKeyserlingk2016} is equivalent to $H^2(\mathbb{Z} \rtimes G, \mathrm{U}(1))$. We do this by exploiting the connection between the second cohomology group and projective representations.

Suppose we have a projective representation $V(g\mathbb{T}^n) : g \in G, n \in \mathbb{Z}$ of $\mathbb{Z} \rtimes G$.
Then we can define a \emph{new} representation $V^{\prime}(g\mathbb{T}^n) = V(g) V(\mathbb{T})^n$. Clearly, since $V$ is a projective representation, $V^{\prime}(g\mathbb{T}^n)$ can differ from $V(g \mathbb{T}^n)$ at most by a phase factor $\chi(g\mathbb{T}^n)$. Thus, defining the corresponding 2-cocycles $\omega$ and $\omega^{\prime}$ by
\begin{equation}
\label{proj_repn}
V(x_1) V(x_2) = \omega(x_1, x_2) V(x_1 x_2),
\end{equation}
where $x_1 = g_1 \mathbb{T}^{n_1}$, etc.\ (and similarly for $\omega^{\prime}$), we find that they are in the same equivalence class [thus, they correspond to the same element of $H^2(\mathbb{Z} \rtimes G, \mathrm{U}(1))$]. On the other hand, $\omega^{\prime}$ is completely determined once we know its restriction $\omega^{\prime}_G$ to $G$ and the extra data $\chi(g) = V(g) \mathbb{T} V(g)^{-1} \mathbb{T}^{-\alpha(g)}$ (where $\alpha(g) = 1$ for unitary $g$ and $-1$ for anti-unitary $g$). One can verify that $\chi$ must satisfy the equation
\begin{equation}
\label{twisted_condn}
\chi(gh) = \chi(g)^{\alpha(h)} \chi(h)^{\alpha(g)}.
\end{equation}
Thus, up to equivalence, the 2-cocycle $\omega$ of $\mathbb{Z} \rtimes G$ is fully determined by a 2-cocycle $\omega^{\prime}_G$ of $G$, and $\chi$ satisfying \eqnref{twisted_condn}. This is indeed the classification of Ref.~\onlinecite{VonKeyserlingk2016}.



\section{Appendix: Deriving the SPT classification}
\label{deriving_spt}
Here we will briefly recap the argument for the $H^{d+1}(G, \mathrm{U}(1))$ classification of SPT ground states in $d = 1$ and $d=2$, taking care to formulate it in such a way as to make it clear that it can also be applied to give a $H^{d+1}(\mathbb{Z} \rtimes G, \mathrm{U}(1))$ classification in Floquet systems.
Suppose we have some short-range entangled state $\ket{\Psi}$ defined on a system without boundary, such that $\ket{\Psi}$ is invariant under the local unitary (or anti-unitary) representation $V(g)$ of a symmetry. Now imagine some subregion $M$ of the whole system, and consider the \emph{subspace} $\mathcal{P}_{M,\ket{\Psi}}$ of ``boundary states'' defined in the Hilbert space of $M$ which \emph{complete} to $\ket{\Psi}$, in the sense that they are identical to $\ket{\Psi}$ away from the boundary of $M$. The restriction $V_M(g)$ of the symmetry operation $V(g)$ to the region $M$ must preserve this subspace (note that this restriction is still well-defined even for anti-unitary symmetries, since we can take it to act only on the Hilbert space of $M$.) Thus, it is well-defined to talk about the action of the symmetry on the boundary states.

Moreover, if we assume that $\ket{\Psi}$ is short-range entangled, this implies that there exists a local unitary $\mathcal{D}$ which transforms $\ket{\Psi}$ into a product state $\ket{\phi}^{\otimes N}$. The restriction $\mathcal{D}_M$ must then transform the states in $\mathcal{P}_{M,\ket{\Psi}}$ into the states which look like a product of $\ket{\psi}$'s away from the boundary. Thus, if we started with a system in $d$ spatial dimensions, we can identify the boundary states with the states of a $(d-1)$-dimensional system. In the case $d=1$, the boundary is just a set of points and we classify the SPT order from the \emph{projective} representation of the symmetry on a boundary point \cite{Pollmann2010,Chen2011,Schuch2011a,Pollmann2012}. In $d=2$, we can classify the SPT order by considering a symmetry restriction procedure as described in Ref.~\onlinecite{Else2014}. 


%
\end{document}